\documentclass[prl,epsfig,twocolumn,showpacs,preprintnumbers,amsmath,amssymb]{revtex4}
\usepackage{amsmath}
\usepackage{epsfig}
\usepackage{graphics}
\begin{document}
\title{Magnetic and superfluid phases of confined fermions in two-dimensional optical lattices}

\author{Brian M.\ Andersen and G.\ M.\ Bruun}

\affiliation{Niels Bohr Institute, University of Copenhagen,
Universitetsparken 5, DK-2100 Copenhagen \O, Denmark.}

\date{\today{}}

\begin{abstract}
We examine antiferromagnetic and $d$-wave superfluid phases of cold
fermionic atoms with repulsive interactions in a two-dimensional
optical lattice combined with a harmonic trapping potential. For
experimentally realistic parameters, the trapping potential leads to
the coexistence of magnetic and superfluid ordered phases with the
normal phase. We study the intriguing shell structures arising from
the competition between the magnetic and superfluid order as a
function of the filling fraction. In certain cases,
antiferromagnetism induces superfluidity by charge redistributions.
We furthermore demonstrate how these shell structures can be
detected  as distinct antibunching dips and pairing peaks in the
density-density correlation function probed in expansion
experiments.

\end{abstract}


\maketitle

The trapping of ultracold atoms in optical lattices opens up the
possibility to study quantum systems in periodic potentials with
unprecedented experimental versatility. One can mimic strongly
correlated systems relevant for paradigmatic condensed matter
applications as well as creating entirely new structures. The pace
of experimental progress is impressive. The  quantum phase
transition between a superfluid and a Mott insulator has been
observed for bosons~\cite{Greiner}, and Fermi surface effects were
reported for fermions~\cite{Esslinger}. Bunching and antibunching
effects in the density-density correlations were found for bosons
and fermions in expansion experiments~\cite{Immanuel} and evidence
of $s$-wave pairing was recently presented~\cite{Ketterle}.

A major goal  is to study two component ($\sigma=\pm$) fermionic
atoms in a two-dimensional (2D) optical lattice with repulsive
interactions. Such a system is well described by the Hubbard model
whose phase diagram is controversial and directly related to the
physics of high-$T_c$ superconductors. For filling fractions  close
to one particle per site ($n=1$), the system is antiferromagnetic
(AF) whereas for smaller densities the true ground state is unknown.
An important question is whether a $d$-wave superconducting state
arises solely from repulsive interactions, but many other open
questions remain, including suggested stripe and checkerboard charge
ordered ground states\cite{Kivelson:2003} that possibly coexist with
superconductivity for some range of doping $x=1-n$. Experimentally,
there is strong evidence for such inhomogeneous states in the
cuprates\cite{Kivelson:2003,Tranquada:1995,Hanaguri:2004}, but it is
unclear if they are important for pairing. An aim of the cold gas
experiments is to improve our understanding of this complicated
problem.

Motivated by this, we study a two component gas of fermionic atoms
in a 2D optical lattice. We include an external harmonic potential
which originates from the Gaussian profile of the laser beams
generating the trap. A main purpose of this paper is to study the
interplay between this harmonic trapping potential and two of the
most dominating ordered phases of the homogeneous Hubbard model: the
antiferromagnetic and $d$-wave superconducting phases. Previous
theoretical studies have focused on the one-dimensional (1D)
case~\cite{rigol}. For realistic parameters, we find that the
magnetic and $d$-wave superfluid (dSF) phases coexist and form shell
structures, in analogy with what has been observed for
bosons~\cite{Folling}. We then examine how these shell structures
can be detected in density-density correlations by expansion
experiments similar to those performed for ideal
gases~\cite{Immanuel}.

A two-component Fermi gas in an optical
lattice is well described by the Hubbard model~\cite{Jaksch,
Hofstetter}
\begin{eqnarray}\nonumber
\hat{H}&=&-t\sum_{\langle ij\rangle,\sigma}\hat{a}_{i\sigma}^\dagger
\hat{a}_{j\sigma}-\mu\sum_{i\sigma}\hat{n}_{i\sigma}+U\sum_i\hat{n}_{i\uparrow}\hat{n}_{i\downarrow}\\
&+&\frac{1}{2}m\omega^2\sum_{i\sigma} R_i^2\hat{n}_{i\sigma}, \label{Hubbard}
\end{eqnarray}
where $\hat{a}_{i\sigma}$ are annihilation operators for localized
atoms on site $i$ [at position ${\mathbf{R}}_i=(X_i,Y_i)$] with spin
$\sigma$ and $\hat{n}_{i\sigma}=\hat{a}_{i\sigma}^\dagger
\hat{a}_{i\sigma}$ is the number operator. The parameters $t$ and
$U>0$ denote hopping between nearest-neighbor sites $\langle
ij\rangle$ and on-site repulsion, respectively. They can be obtained
from the atom-atom scattering length and the lowest band Wannier
state~\cite{Zwerger}. The last term in Eq.\eqref{Hubbard} describes
the harmonic trapping potential. We use unrestricted Hartree-Fock
approximation to decouple the interaction term in
Eq.\eqref{Hubbard}.

In addition to the AF state at half-filling, numerical methods
indicate that the repulsive Hubbard model prefers $d$-wave
superconducting order in a certain region of the $U-x$ phase
diagram\cite{dwaveinHub}. Therefore, in a harmonic trap we expect
that AF and superfluid order may coexist in the atomic cloud. To
model this we include explicitly a BCS $d$-wave term so that the
final mean-field Hamiltonian becomes
\begin{gather}\label{HubbardMF}
\hat{H}=-t\sum_{\langle ij\rangle,\sigma}\hat{a}_{i\sigma}^\dagger
\hat{a}_{j\sigma}-\mu\sum_{i\sigma}\hat{n}_{i\sigma}+\frac{1}{2}m\omega^2\sum_{i\sigma} R_i^2\hat{n}_{i\sigma}\\
+U\sum_i \left[ \langle \hat{n}_{i\uparrow}\rangle
\hat{n}_{i\downarrow} + \hat{n}_{i\uparrow}\langle
\hat{n}_{i\downarrow}\rangle \right] + \sum_{\langle ij \rangle}
\left[ \Delta_{ij} a^\dagger_{i\uparrow} a^\dagger_{j\downarrow} +
\mbox{H.c} \right].\nonumber
\end{gather}
It is well-known that a nearest-neighbor attraction can generate
such a $d$-wave BCS term at the self-consistent
level\cite{allHamiltonian}. We stress that even though our
mean-field model Eq.\eqref{HubbardMF} is phenomenological, it
captures the existence and competition of ordered phases at $T=0$
such as AF and superfluidity, and has been widely used previously in
the high-$T_c$ community\cite{allHamiltonian}. Note that even in 1D
it is possible to obtain qualitative information from mean-field
theory, such as e.g. the shape of the density profiles and the
existence of antiferromagnetic correlations\cite{rigol}. This gives
further confidence in the 2D $T=0$ mean-field results presented
below.

Equation \eqref{HubbardMF} can be diagonalized by the
transformation, ${\hat{a}_{i\sigma}}^\dagger = \sum_{E_{n\sigma}>0}
[u^*_{n\sigma}(i) \hat{\gamma}_{n\sigma}^\dagger + \sigma
v_{n\sigma}(i) {\hat{\gamma}}_{n-\sigma}]$ yielding the  Bogoliubov-de Gennes equations
\begin{equation}\label{BdG}
\sum_j \left( \begin{array}{cc} {\mathcal{K}}^{+}_{ij}& {\mathcal{D}}_{ij} \\
{\mathcal{D}}^*_{ij} & -{\mathcal{K}}^{-}_{ij}
\end{array} \right) \left( \begin{array}{c} u_{n\sigma}(j) \\ v_{n-\sigma}(j)
\end{array}
\right) = E_{n\sigma} \left( \begin{array}{c} u_{n\sigma}(i) \\
v_{n-\sigma}(i) \end{array} \right).
\end{equation}
The diagonal blocks are given by
${\mathcal{K}}^{\pm}_{ij}=-t\delta_{\langle
ij\rangle}+(V_i-\mu+U\langle
\hat{n}_{i\mp\sigma}\rangle)\delta_{ij}$,  where
$V_i=\frac{1}{2}m\omega^2 R_i^2$, and $\delta_{ij}$ and
$\delta_{\langle ij \rangle}$ are the Kronecker delta symbols
connecting on-site and nearest neighbor sites, respectively. The
off-diagonal block is ${\mathcal{D}}_{ij}=\delta_{\langle ij
\rangle} \Delta_{ij}$, where in the homogeneous case
$\Delta_{ij}=+(-)\Delta$ on the $x$($y$)-links corresponding to bulk
$d_{x^2-y^2}$-wave pairing symmetry. Below, we restrict the
discussion to $T=0$ and enforce self-consistency through iteration
of the relations $\langle \hat{n}_{i\sigma} \rangle = \sum_{n}
|v_{n\sigma}(i)|^2$ and $\Delta_{ij}=V_d \langle
\hat{a}_{i\uparrow}\hat{a}_{j\downarrow}-\hat{a}_{i\downarrow}\hat{a}_{j\uparrow}
\rangle=V_d \sum_{n} [ u_{n\uparrow}(i) v_{n\downarrow}^*(j) -
u_{n\uparrow}(j) v_{n\downarrow}^*(i)]$. Here, $V_d$ is a coupling
constant which, in principle, is a function of $U$, but at the
phenomenological level becomes an independent parameter.

We now present our results of the numerical solution of the
mean-field equations varying the number of particles $N$ trapped in
a potential with $\frac{1}{2}m\omega^2d^2/t=0.025$ where $d=1$ is
the lattice spacing. This yields experimentally realistic lattice
sizes of the order $\sim40\times 40$. For these relatively small
systems and the parameters used in this paper, the superfluid
coherence length $\xi$ is comparable to the characteristic length
scale $\lambda$ of the ordered phases. Therefore, we include the
trapping potential exactly since the local density approximation
cannot be expected to be valid because it assumes that $\xi \ll
\lambda$.

\begin{figure}[t]
\includegraphics[clip=true,width=.95\columnwidth,height=10cm]{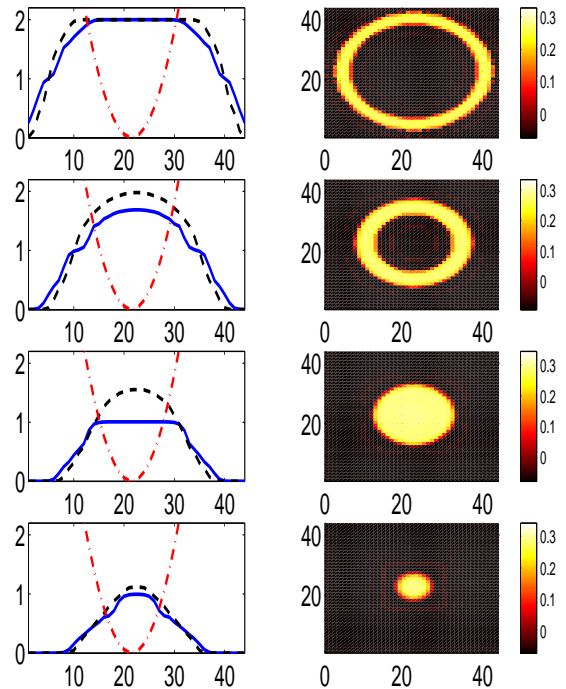}
\caption{(Color online) Left: line cut of the density (blue, solid
lines) and trap potential (red, dot-dashed lines) through the center
of the (spherically symmetric) trap, $U/t=4.0$. The black dashed
lines show the density profile for the noninteracting case, $U=0$.
Right: the associated real-space staggered magnetization. The total
number of fermions in the $44\times 44$ system are (top to bottom):
1936, 968, 484, 242.} \label{fig:densmagnU4Vd0}
\end{figure}

We first discuss the situation with no dSF order, i.e. $V_d=0$, and
use an $N\times N$ square lattice ($N=44$) with open boundary
conditions. In Fig. \ref{fig:densmagnU4Vd0} we plot the density
profile $n_i=\langle \hat{n}_{i\uparrow} \rangle + \langle
\hat{n}_{i\downarrow} \rangle$ and the staggered magnetization
$m_i=(-1)^{X_i+Y_i}(\langle \hat{n}_{i\uparrow} \rangle - \langle
\hat{n}_{i\downarrow} \rangle)/2$ with $U/t=4.0$ for a varying
number of trapped particles. The dashed lines in the left column
display the density profile for an ideal gas ($U=0$). For
sufficiently high density (top panel), the center region is a band
insulator in both the interacting and non-interacting limits. We see
that the main effect of the interaction is the formation of AF
regions at densities $n\simeq1$. Since the system is inhomogeneous
due to the trapping potential, the AF order coexists with the normal
phase. This  leads to steps in the density profile as the magnetic
correlations favor $n\simeq1$. At the same time, the density is
reduced in the center of the trap and the atomic cloud becomes more
extended. Upon reducing the number of particles in the trap, the
magnetization is seen to evolve from a ring structure to a center
island.

For larger systems, spin-density waves with ordering vectors other
than the conventional AF ${\mathbf{Q}}=(\pi,\pi)$, e.g. stripe
phases known from high-T$_c$ materials, may become evident at
filling away from $n\simeq1$. In that case we expect the same
overall results as those in Fig. \ref{fig:densmagnU4Vd0} but with
magnetism existing for a wider doping range.

\begin{figure}[t]
\includegraphics[clip=true,width=.95\columnwidth,height=10cm]{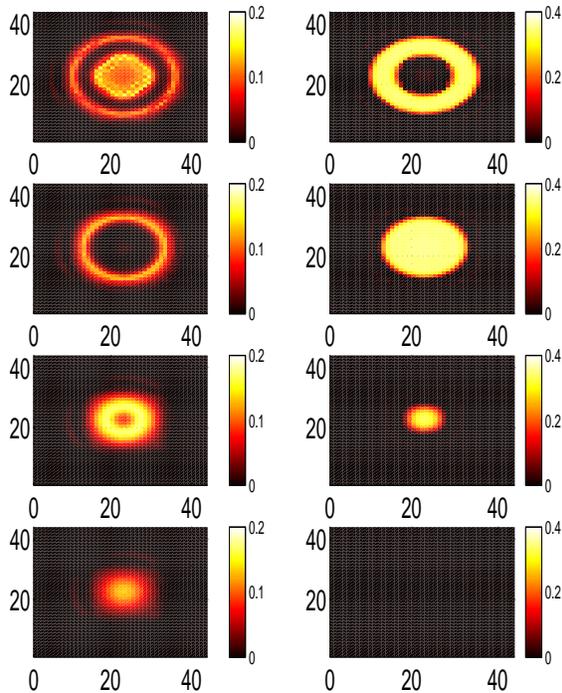}
\caption{(Color online) Left: spatial dependence of the $d$-wave
superfluid order $\Delta_i$ with $V_d/t=2.0$ and $U/t=4.0$. Right:
staggered magnetization obtained for the same set of parameters. The
total number of fermions are (top to bottom): 726, 484, 242, 144.}
\label{fig:densmagnU4Vd2}
\end{figure}

Next we discuss the possibility of superfluid order focusing on both
the spatial distribution and the interplay with the magnetic order.
In Fig. \ref{fig:densmagnU4Vd2} we show the $d$-wave order parameter
(left) defined as
$\Delta_i=(1/4)(\Delta_{i,i+\hat{x}}+\Delta_{i,i-\hat{x}}-\Delta_{i,i+\hat{y}}-\Delta_{i,i-\hat{y}})$,
where $\hat{x}(\hat{y})$ are the unit vectors along the $x(y)$ axis,
and the magnetization (right column). The density profiles (not
shown) are similar to those presented in Fig.
\ref{fig:densmagnU4Vd0}. The spatial structures depicted in
Fig.\ref{fig:densmagnU4Vd2} come from the interplay between the
trapping potential and the magnetic and superfluid correlations and
can be understood as follows. The amplitude of $\Delta$ peaks at one
particle per site just like the AF order. This leads to a
competition between dSF and AF order in regions around $n\simeq 1$.
We have chosen $V_d/t=2.0$ giving $\Delta\sim0.15t$ (and a coherence
length of a few lattice spacings) consistent with the numerical
results of Ref. \onlinecite{dwaveinHub}. For the resulting ratio
$V_d/U=1/2$, antiferromagnetism dominates and dSF order is generally
left to exist in regions surrounding the AF order. This is seen, for
example, in the two middle rows of Fig. \ref{fig:densmagnU4Vd2}
where AF exists in a center island with $n\simeq1$ surrounded by a
ring of dSF. This is also the origin of the outer rim of the
remarkable double-ring structure shown in the top left subplot of
Fig. \ref{fig:densmagnU4Vd2}. There, however, the center dSF island
is a case where magnetism surprisingly promotes dSF because of
charge redistributions: The AF order decreases the density in the
center of the trap below the threshold for generating dSF. For
sufficiently small fillings, only a superfluid cloud is left in the
trap (bottom row in Fig. \ref{fig:densmagnU4Vd2}).

We now address the important question of how the shell structures
presented in Figs. \ref{fig:densmagnU4Vd0} and
\ref{fig:densmagnU4Vd2} can be detected experimentally. Since
magnetic and superfluid order coexist with the normal phase, it is
{\it a priori} unclear how strong their experimental signatures will
be. A well known experimental technique is  to measure the
density-density correlations of an expanding gas after the lattice
has been switched off. The density-density correlation function
$\langle n({\mathbf{r}})n({\mathbf{r}}')\rangle$ after expansion
time $t$ probes the momentum correlation function $\langle
n_{\mathbf{q}}n_{\mathbf{q}'}\rangle$ for the interacting system
before the expansion with $ n_{\mathbf{q}}= \sum_\sigma
a_{{\mathbf{q}}\sigma}^\dagger a_{{\mathbf{q}}\sigma}$ and
${\mathbf{r}}={\mathbf{q}}t/m$~\cite{Altman}. We focus on the
correlation function
\begin{eqnarray}\label{corr}
{\mathcal{C}}({\mathbf{q}},{\mathbf{q}'})=\langle
n_{\mathbf{q}}n_{\mathbf{q}'}\rangle&-&\langle n_{\mathbf{q}}\rangle
\langle n_{\mathbf{q}'}\rangle=
\delta_{{\mathbf{q}},{\mathbf{q}'}}\langle n_{\mathbf{q}}
\rangle\\\nonumber - |\sum_{n\sigma} b^*_{{\mathbf{q}'}n\sigma}
b_{{\mathbf{q}}n\sigma}|^2&+&\sum_{nm\sigma}
b^*_{{\mathbf{q}'}n\sigma} a^*_{{\mathbf{q}}n\sigma}
a_{{\mathbf{q}}m\sigma} b_{{\mathbf{q}'}m\sigma},
\end{eqnarray}
where $a_{{\mathbf{q}}n\sigma}(b_{{\mathbf{q}}n\sigma})=(1/N) \sum_i
u_{in\sigma} (v_{in\sigma}) \exp (-i {\mathbf{q}} \cdot
{\mathbf{R}_i})$. For non-interacting fermions,
${\mathcal{C}}({\mathbf{q}},{\mathbf{q}'})$ has antibunching dips
for ${\mathbf{q}}-{\mathbf{q}'}=(n_x2\pi,n_y2\pi)$ where $n_x$ and
$n_y$ are integers  as was recently observed~\cite{Immanuel}. Atoms
forming an AF state will in addition exhibit antibunching dips in
${\mathcal{C}}({\mathbf{q}},{\mathbf{q}'})$ for
${\mathbf{q}}-{\mathbf{q}'}=(n_x\pi,n_y\pi)$ with odd integers
$n_x,n_y$, reflecting the period doubling due to the magnetic order.
For a system without an external trapping potential characterized by
an AF order parameter $m=|\langle n_{i\uparrow}\rangle-\langle
n_{i\downarrow}\rangle|/2$, mean-field theory yields $ \langle
n_{\mathbf{q}}n_{\mathbf{q}'}\rangle=\langle
n_{\mathbf{q}}\rangle\langle n_{\mathbf{q}'}\rangle-
\sum_\sigma\langle a_{{\mathbf{q}}\sigma}^\dagger
a_{{\mathbf{q}}'\sigma}\rangle \langle
a_{{\mathbf{q}}'\sigma}^\dagger a_{{\mathbf{q}}\sigma}\rangle $ with
$\langle a_{{\mathbf{q}}\sigma}^\dagger
a_{{\mathbf{q}}'\sigma}\rangle=(Um/2)/2E_q$ for
${\mathbf{q}}=\mathbf{q}'+(\pi,\pi)$ and
$E_q=\sqrt{\epsilon_q^2+(Um/2)^2}$. Here,
$\epsilon_q=-2t[\cos(q_x)+\cos(q_y)]$ is the usual tight-binding
spectrum. Likewise, for a homogeneous superfluid $d$-wave state, BCS
theory yields $\langle n_{\mathbf{q}}n_{\mathbf{q}'}\rangle=\langle
n_{\mathbf{q}}\rangle\langle n_{\mathbf{q}'}\rangle+
\sum_\sigma|\langle a_{{\mathbf{q}}\sigma}
a_{{\mathbf{q}}'-\sigma}\rangle |^2$ with $|\langle
a_{{\mathbf{q}}\sigma}
a_{-{\mathbf{q}}-\sigma}\rangle|=|\Delta_q|/2E_q$, for
${\mathbf{q}}=-\mathbf{q}'$ and
$E_q=\sqrt{(\epsilon_q-\mu)^2+\Delta_q^2}$ where
$\Delta_q=2\Delta[\cos(q_x)-\cos(q_y)]$ is the $d$-wave gap. Pair
correlations with $s$-wave symmetry were measured on the
Bose-Einstein condensate side of a homogeneous system with a
Feshbach resonance~\cite{Greiner2}.

\begin{figure}[t]
\begin{center}
\leavevmode
\begin{minipage}{.49\columnwidth}
\includegraphics[clip=true,width=.99\columnwidth]{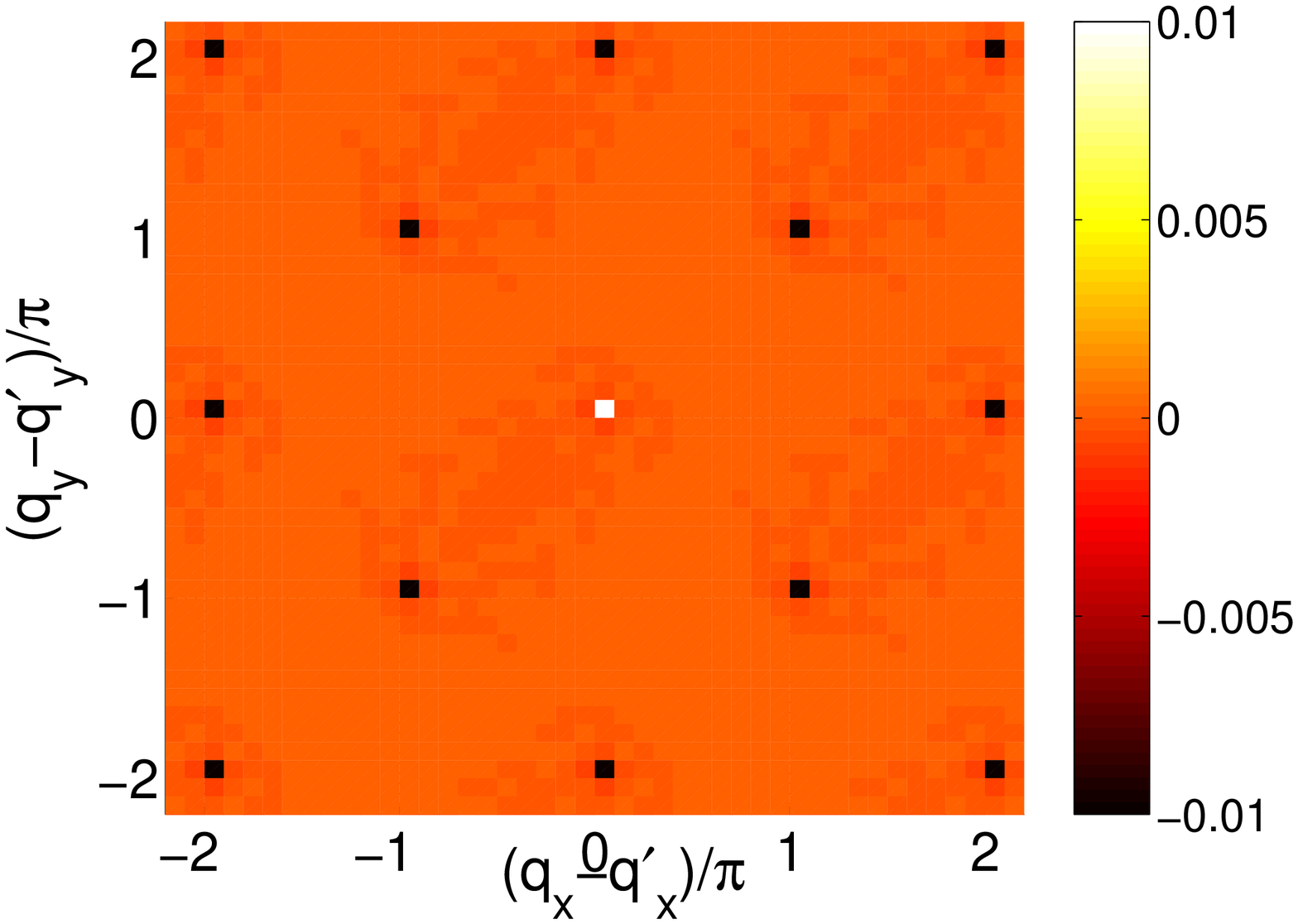}
\end{minipage}
\begin{minipage}{.49\columnwidth}
\includegraphics[clip=true,width=.99\columnwidth]{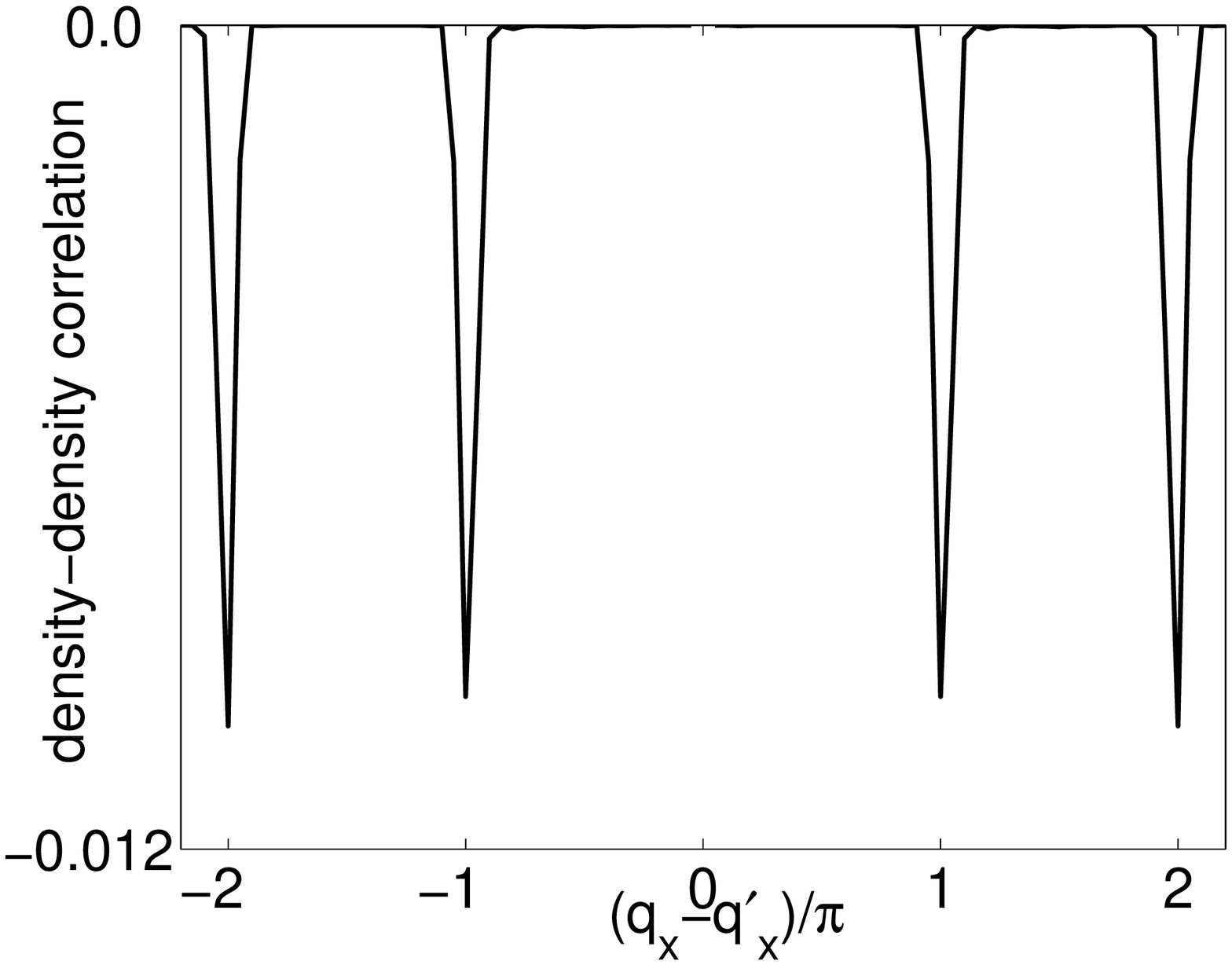}
\end{minipage}\\
\begin{minipage}{.49\columnwidth}
\includegraphics[clip=true,width=.99\columnwidth]{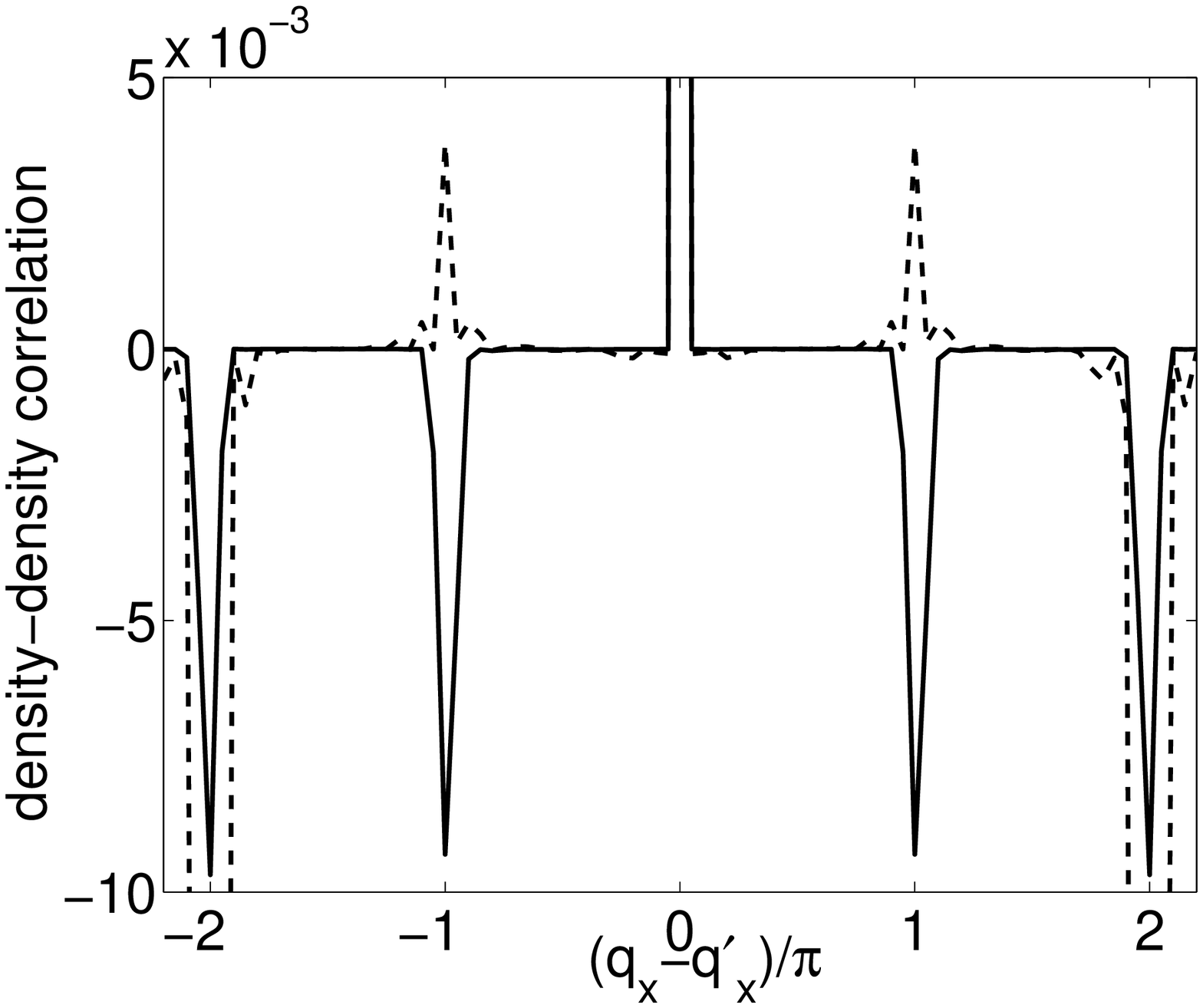}
\end{minipage}
\begin{minipage}{.49\columnwidth}
\includegraphics[clip=true,width=.99\columnwidth]{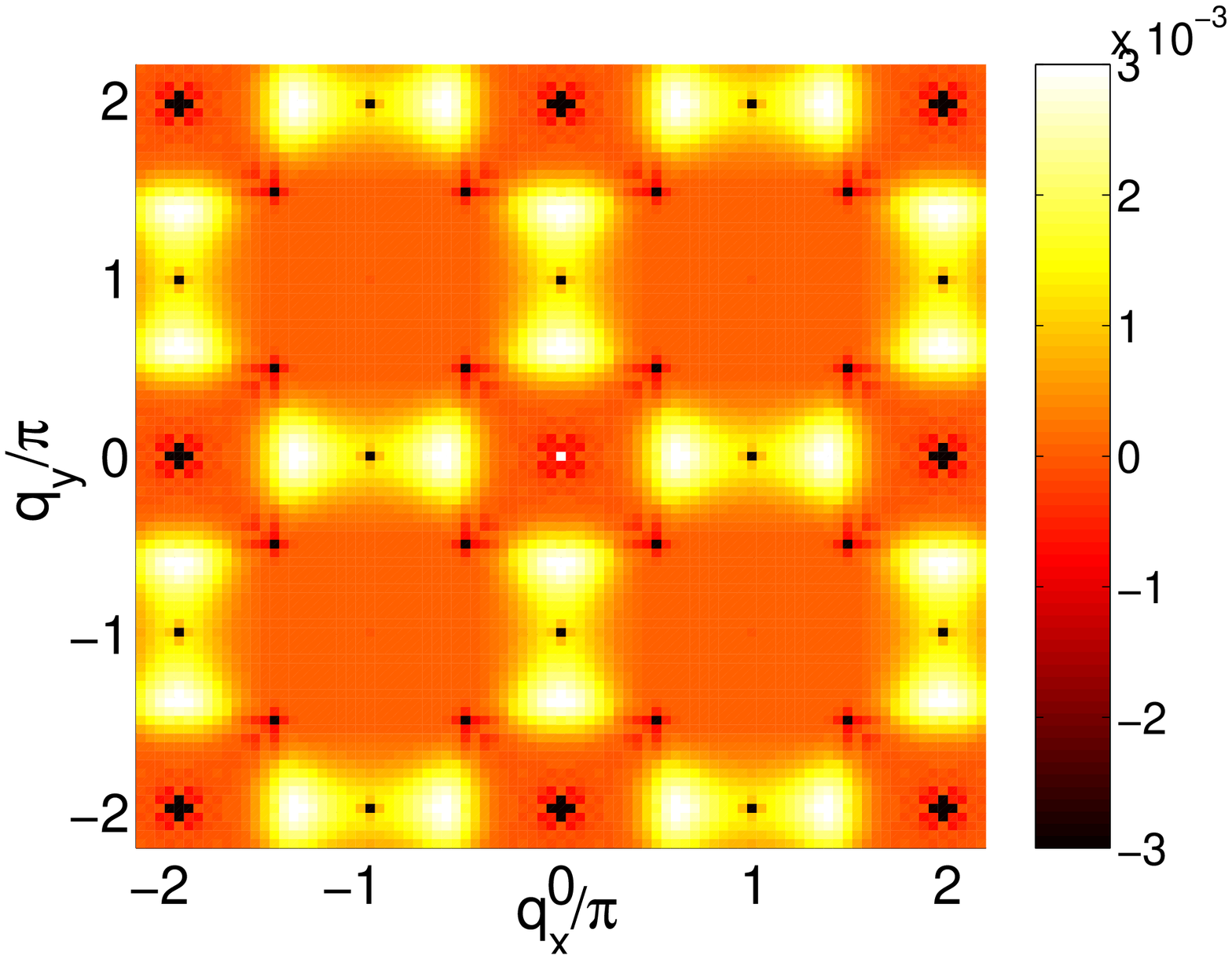}
\end{minipage}
\caption{(Color online) (top row) Density-density correlation
function ${\mathcal{C}}$ versus ${\mathbf{q}}-{\mathbf{q}'}$ (left)
with ${\mathbf{q}'}=(\pi/2,\pi/2)$, and a cut (right) along the
diagonal $q_x-q_x'=q_y-q_y'$. Parameters are identical to those used
in Fig. \ref{fig:densmagnU4Vd0} (third row). (lower left)
${\mathcal{C}}({\mathbf{q}},{\mathbf{q}'})$ along a diagonal with
$q_x-q_x'=q_y-q_y'$ and ${\mathbf{q}'}=(\pi/2,\pi/2)$ (solid line)
and along a horizontal cut $q_x-q_x'$ with $q_y=0$ and
${\mathbf{q}'}=(\pi/2,0)$ (dashed line) for the same parameters as
in Fig. \ref{fig:densmagnU4Vd2} (second row). (lower right)
${\mathcal{C}}({\mathbf{q}},{\mathbf{q}'})$ versus ${\mathbf{q}}$ when
${\mathbf{q}'}=-{\mathbf{q}}$.}\label{fig:antibunch}
\end{center}
\end{figure}

In order to examine how the antibunching dips  and the pairing
correlation peaks show up for the trapped lattice, we calculate
${\mathcal{C}}({\mathbf{q}},{\mathbf{q}'})$ for the shell structures
shown in Fig. \ref{fig:densmagnU4Vd0} and \ref{fig:densmagnU4Vd2}.
In Fig. \ref{fig:antibunch}(top left), we show a 2D map of
${\mathcal{C}}({\mathbf{q}},{\mathbf{q}'})$ with
${\mathbf{q}'}=(\pi/2,\pi/2)$ for the same parameters used in Fig.
\ref{fig:densmagnU4Vd0} (third row). Figure \ref{fig:antibunch}(top
right) shows a cut along the diagonal $q_x-q_x'=q_y-q_y'$. Here, in
addition to the usual lattice dips\cite{Immanuel}, we clearly see
the signature of the AF state in the additional antibunching dips
at $q_x-q_x'=(\pm\pi,\pm\pi)$. The 
 momentum 
${\mathbf{q}'}=(\pi/2,\pi/2)$ is close to the Fermi surface and
hence maximizes the ratio of the AF dips to the $2\pi$ lattice dips.
Note that the perfect periodicity, i.e. the equal amplitude at 
points ${\mathbf{q}}\rightarrow {\mathbf{q}}+(n_x2\pi,n_y2\pi)$, is
an artifact of the model which neglects the spatial extend of the
lowest Wannier function.

We next discuss ${\mathcal{C}}({\mathbf{q}},{\mathbf{q}'})$ with
both AF and dSF order present, and focus on the same
particle filling as above which corresponds to the second row in
Fig. \ref{fig:densmagnU4Vd2}. We show in Fig.
\ref{fig:antibunch}(bottom left) momentum cuts in
${\mathcal{C}}({\mathbf{q}},{\mathbf{q}'})$ along a diagonal line
with $q_x-q_x'=q_y-q_y'$ and ${\mathbf{q}'}=(\pi/2,\pi/2)$ (solid
line) and along a horizontal line $q_x-q_x'$ with $q_y=0$ and
${\mathbf{q}'}=(\pi/2,0)$ (dashed line). Here, in addition to the AF
dips, it is evident that the dSF phase displays bunching at
${\mathbf{q}}=-{\mathbf{q}'}$ since the peaks appear at
$q_x-q_x'=q_x-\pi/2=-\pi$ and $\pm2\pi$ displacements thereof. In
Fig. \ref{fig:antibunch}(bottom right) we fix
${\mathbf{q}}=-{\mathbf{q}'}$ and plot a 2D map of
${\mathcal{C}}({\mathbf{q}},-{\mathbf{q}})$.
This clearly reveals the $d$-wave symmetry of the pairing: the
 bunching is maximum along the $x$ and $y$ directions and minimum
along  $x=\pm y$. Also, the pairing is maximum around the Fermi surface
as expected.

The dSF pairing peaks in Fig. \ref{fig:antibunch} are small and may
be easier to detect at lower filling fractions where AF order is
absent and a larger fraction of the particles participate in the
pairing. Note that the AF antibunching dips are also present in Fig.
\ref{fig:antibunch}(bottom right) at
${\mathbf{q}}=-{\mathbf{q}'}=(\pm\pi/2,\pm\pi/2)$. This is an
example of how the co-existence of the magnetic and superfluid order
can be detected as the presence of both AF antibunching and dSF
bunching peaks in expansion experiments. The AF dips disappear for
lower fillings when the Fermi surface moves below
$(\pm\pi/2,\pm\pi/2)$ and the density vanishes there. For the same
reason there are no lattice dips at $(\pm\pi,\pm\pi)$ in this image.

In summary, we have studied the magnetic and superfluid phases of
repulsive fermionic atoms on a 2D lattice combined with a harmonic
trap for experimentally realistic parameters. The Hubbard
correlations result in intriguing magnetic real-space shell
structures which may coexist and compete with a superfluid phase.
These phases show up as distinct antiferromagnetic antibunching dips
and superfluid bunching peaks with $d$-wave symmetry in the
density-density correlations which can be probed in expansion
experiments. The results are relevant to current experiments on
atoms in optical lattices.

B. M. A. is supported by the Danish Technical Research Council via
the Framework Programme on Superconductivity.

\end{document}